\documentclass[reprint, aps, prl, longbibliography, notitlepage]{revtex4-2}
\usepackage{gensymb}
\usepackage{amsmath, amssymb}
\usepackage{graphicx}
\usepackage{float} 
\usepackage{physics}
\usepackage{soul}
\usepackage[dvipsnames]{xcolor}
\usepackage[colorlinks, linkcolor= blue, citecolor = blue, urlcolor=blue]{hyperref}
\usepackage{comment}
\usepackage{pifont} 
\usepackage{multirow}
\usepackage[version=4]{mhchem}
\usepackage[normalem]{ulem}

\def\be{\begin{equation}}
\def\ee{\end{equation}}
\def \bea{\begin{eqnarray}}
\def \eea{\end{eqnarray}}

\def\kb{\mathbf k}
\def\bm{\mathbf}

\begin{document}
\title{Spin band geometry drives intrinsic thermal spin magnetization and current}

\author{Sankar Sarkar}
\affiliation{Department of Physics, Indian Institute of Technology Kanpur, Kanpur-208016, India}
\author{Harsh Varshney}
\affiliation{Department of Physics, Indian Institute of Technology Kanpur, Kanpur-208016, India}
\author{Sayan Sarkar}
\affiliation{Department of Physics, Indian Institute of Technology Kanpur, Kanpur-208016, India}
\author{Amit Agarwal}
\email{amitag@iitk.ac.in}
\affiliation{Department of Physics, Indian Institute of Technology Kanpur, Kanpur-208016, India}

\begin{abstract}
Generating spin magnetization and spin currents without magnetic or electric fields is a key frontier in spin caloritronics. Spin responses driven by thermal gradients offer a promising route, though the band geometric origin of intrinsic mechanisms, especially in non-magnetic materials, remains poorly understood. Here we develop a unified quantum theory of thermal spin magnetization and spin currents in itinerant electrons, rooted in spin band geometry with both Fermi-surface and Fermi-sea contributions. We identify two key geometric quantities: the spin-velocity metric tensor, which governs thermal spin magnetization, and the spin geometric tensor, combining spin Berry curvature and spin quantum metric, which generates thermal spin currents. These intrinsic contributions persist and can even dominate in non-magnetic insulators. Numerical calculations for chiral metal RhGe and antiferromagnet CuMnAs demonstrate sizable thermal spin responses near band crossings. Our results establish the band geometric origin of thermal spin transport and provide guiding principles for discovering and engineering next-generation spin caloritronic materials.
\end{abstract}

\maketitle

\textcolor{blue}{\it Introduction:--} %
The ability to generate and control spin magnetization and spin currents without relying on magnetic fields is a central challenge for next-generation spintronic technologies~\cite{Igor_2004, Bader_2010, qiu_2014, yoon_2017, Manchon_2019,Ryu_2020, shao_2021, kim_2024}. 
Thermal gradients, naturally present in micro- and nanoscale devices, provide a compelling route to drive such spin responses in an energy-efficient manner compatible with device miniaturization. 
Thermally induced spin responses, such as thermal spin magnetization (TSM)\cite{WANG_2010, Dyrdal_2013, xiao_2016, Dyrdal_2018} and thermal spin currents (TSC)\cite{uchida_2008, jaworski_2010, uchida_2010, Tauber_2012, sheng_2017, meyer_2017}, enable heat-to-spin conversion and promise low-power spin-caloritronic applications~\cite{Bauer_2012, Yu_2017}, with implications for high-speed, energy-efficient memory and computing devices~\cite{ geranton_2015, freimuth_2016, Kim_ACSnano2020, shao_2021}. 

Most prior studies have focused on current-induced spin magnetization in inversion-broken systems~\cite{V.M.Edelstein_1990, Kato_2004_EISP, Gorini_2017} and on spin Hall currents in non-magnetic systems preserving time-reversal symmetry $\mathcal{T}$~\cite{Hirsch_1999, Kato_2004, Sinova_2015}. In contrast, thermally generated spin responses have received comparatively less attention. Existing studies of spin-caloritronic effects~\cite{Cheng_2016, Zyuzin_2016, Rezende_2016} emphasize magnon-mediated transport in ordered magnets, while corresponding mechanisms for Bloch electrons in non-magnetic systems remain unexplored. By comparison, charge transport is firmly understood as a band-geometric phenomenon governed by Berry curvature and quantum metric~\cite{Xiao_2010, Nagaosa_2010, Onoda_2006, Bhalla_2022, kamal_2023, harsh, harsh2, Debottam_2024, Maneesh_2024, varshney_2024, adak_2024}. Yet, the role of band geometry in thermal spin transport has not been clarified. Although recent works have linked quantum geometry to current-induced spin responses, a parallel framework for thermal driving, essential for harnessing waste heat in non-magnetic platforms, remains absent. This gap limits progress toward next-generation spin-caloritronic materials and devices and motivates the central question of this work: {\it Can the spin band geometry of Bloch electrons drive robust thermal spin magnetization and currents in systems without net magnetic order?}

In this Letter, we develop a unified quantum theory of thermally induced spin transport using the density-matrix formalism. We show that intrinsic (scattering–independent) TSM and TSC are governed by two spin-band-geometric quantities: the spin-velocity metric tensor, which controls TSM, and the spin geometric tensor, combining spin Berry curvature and spin quantum metric~\cite{spin_berry_curvature1, spin_berry_curvature2, spin_berry_curvature4, spin_berry_curvature5, spin_berry_curvature3}, which governs TSC. We also identify extrinsic scattering-dependent contributions arising from band velocity and anomalous spin velocity. A comprehensive symmetry analysis decomposes these responses into $\mathcal{T}$-even and $\mathcal{T}$-odd parts, providing a complete classification of the TSM and TSC responses across all 122 magnetic point groups. Our first-principles and model calculations confirm sizable TSM and TSC in $\mathcal{T}$-symmetric RhGe and $\mathcal{T}$-broken CuMnAs, which are strongly enhanced near the band edges and Dirac crossings. Together, these results establish a band-geometric framework for thermal spin transport and offer symmetry-guided design principles for discovering next-generation spin-caloritronic materials. 

\begin{figure}
    \centering
    \includegraphics[width= 0.9\linewidth]{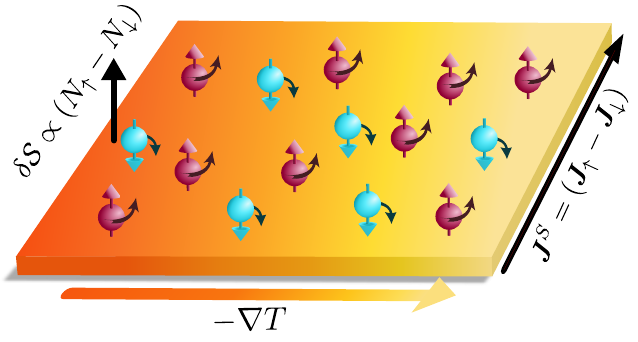}
    \caption{\textbf{Schematic of thermally driven spin responses:} A temperature gradient generates an imbalance between up and down spins of itinerant electrons due to spin-orbit coupling, leading to spin magnetization ($\delta\mathcal{S}$) and  current (${J}^S$).
    \label{fig_1}}
\end{figure}
\textcolor{blue}{\it Thermal spin magnetization and current:--}
A uniform temperature gradient can induce steady-state spin magnetization and current in itinerant electrons, even in non-magnetic materials. To capture these responses quantitatively, we define the spin polarization and spin current as \cite{spin-current1, spin-current2, spin-current3, spin-current5, Manchon_2019, sayan, Sayan_2025}
\begin{subequations}\label{eq:TSM_and_TSC_def}
\begin{align}
    \delta\mathcal{S}^{\gamma}(\omega) &= \langle \hat{s}^\gamma\rangle \equiv \sum_{\kb} {\rm Tr}[~\hat{s}^{\gamma}~~\hat{\rho}~]~, 
\end{align}
\begin{align}
    J^{\gamma; a}(\omega) &= \langle \hat{\mathcal J}^{\gamma; a} \rangle \equiv  \sum_{\kb}\text{Tr}[\hat{\mathcal{J}}^{\gamma; a}~\hat{\rho}]~.
\end{align}
\end{subequations}
Here, $\hat{s}^\gamma$ denotes the operator for spins, polarized along $\gamma \in (x, y, z)$, while  the spin current operator $\hat{\mathcal{J}}^{\gamma; a} = (1/2)\{\hat{v}^a,s^{\gamma}\}$ describes the flow of spins, polarized along $\gamma$, in the $a$-direction with velocity $\hat{v}$. The nonequilibrium density matrix $\hat{\rho} (\mathbf{k}, t)$ encodes the system's response to a harmonically varying time-dependent thermal gradient of frequency $\omega$~\cite{Liu_NN2025,Hirata_APL2025, varshney_2024},
\(
\mathbf E_{T}(t)=-(\nabla T/T)\,[e^{i\omega t}+e^{-i\omega t}]/2~.
\)
We treat this thermal field as a Luttinger gauge field~\cite{Luttinger,tatara,boyd,Sekine,Culcer,harsh,harsh2,Bhalla_2023,Sankar_2025} and calculate $\hat{\rho} (\mathbf{k}, t)$ using the quantum Liouville equation (see Sec. \textcolor{blue}{S1} of Supplementary material (SM) \cite{SM}). 

In linear response, we express TSM and TSC as 
\begin{subequations}
    \begin{align}
        \delta\mathcal S^{\gamma}(\omega)
        &= \frac{\nabla_aT}{T}\,
          \bigl[\chi^{\gamma;a}(\omega)\,e^{-i\omega t}+{\rm c.c.}\bigr]~, \label{eq:TSM_def}\\ 
         J^{\gamma;a}(\omega) &=\frac{\nabla_bT}{T}\!
        \Bigl[\sigma^{\gamma;ab}(\omega)\,e^{-i\omega t}+{\rm c.c.}\Bigr]~, \label{eq:TSC_def}
    \end{align}
\end{subequations}
where $\chi^{\gamma;a}(\omega)$ is the TSM susceptibility tensor and $\sigma^{\gamma;ab}(\omega)$ is the TSC conductivity tensor. Here, the repeated indices are summed over. We calculate TSM and TSC using Eq.~\eqref{eq:TSM_and_TSC_def} up to linear order in ${\bm E}_T$, as discussed in Sec.~\textcolor{blue}{S2} of SM~\cite{SM}. The TSM susceptibility naturally decomposes as  
$\chi^{\gamma;a} = \chi^{\gamma;a}_{\mathrm{Surf}} + \chi^{\gamma;a}_{\mathrm{Sea}}$, with 
\begin{align} \label{eq:TSM_susc}
\chi^{\gamma;a}_{\mathrm{Surf}}(\omega) &=
\frac{1}{2}\sum_{n,\mathbf k}\;
        \tilde\varepsilon_{n\mathbf k}\,
        s^{\gamma}_{nn}\, 
        v^{a}_{nn}\,
        \frac{\partial f_n}{\partial \varepsilon_{n \mathbf{k}}}\,
        \frac{1}{1/\tau-i\omega}~, \\[4pt]
\chi^{\gamma;a}_{\mathrm{Sea}}(\omega) &=
\frac{1}{2\hbar}\sum_{n\neq p,\mathbf k}
        \mathcal Q^{\gamma;a}_{np}
        \frac{(\tilde\varepsilon_{n\mathbf{k}}f_n-\tilde\varepsilon_{p\mathbf{k}}f_p)}
             {\omega_{np}}
        \frac{1}{\frac{1}{\tau}-i\!\left[\omega-\omega_{np}\right]}~. \nonumber 
\end{align}
Here,   
\(\omega_{np}=(\varepsilon_{n\mathbf{k}}-\varepsilon_{p\mathbf{k}})/\hbar\),  
\(v^{a}_{nn}=(1/\hbar)\,\partial\varepsilon_{n\mathbf{k}}/\partial k_{a}\),  
and \(\tilde\varepsilon_{n}=\varepsilon_{n\mathbf{k}}-\mu\). This decomposition highlights two distinct geometric origins: the intraband Fermi-surface term and the interband Fermi-sea term. The interband term, $\chi_{\mathrm{Sea}}$, reflects a Fermi-sea contribution that survives even in insulators. It is driven by a unique band geometric quantity, the \emph{spin-velocity metric tensor},
\begin{equation}
\mathcal Q^{\gamma;a}_{np}=v^{a}_{np}s^{\gamma}_{pn} = \mathcal V^{\gamma;a}_{np}-\tfrac{i}{2}\mathcal U^{\gamma;a}_{np}~,
\end{equation}
with $n\neq p$. Here, the real part \(\mathcal V\) is the anomalous spin velocity and the imaginary part \(\mathcal U\) is related to the anomalous spin polarizability~\cite{Guo_24,Xiang_2025}. A central result of this work is the expression for the Fermi-sea 
contribution $\chi_{\mathrm{Sea}}$ governed by these novel band-geometric quantities. In contrast, the intraband term  
\(\chi_{\mathrm{Surf}}\) arises solely from the Fermi surface and it vanishes in insulating systems \cite{Dyrdal_2013,Dyrdal_2018}. 

\begin{table}[t!]
\caption{\textbf{Thermal spin responses:} 
Decomposition of TSM susceptibility ($\chi^{\gamma;a}$) and thermal spin conductivity ($\sigma^{\gamma;ab}$), given in Eqs.~(\ref{eq:TSM_susc}), (\ref{eq:TSC_surf}), and (\ref{eq:TSC_sea}), into momentum-resolved contributions $\tilde{\chi}^{\gamma;a}(\mathbf{k})$ and $\tilde{\sigma}^{\gamma;ab}(\mathbf{k})$ (sum over relevant bands assumed). The table separates these into Fermi-surface and Fermi-sea terms, and further into $\mathcal{T}$-even and $\mathcal{T}$-odd parts (labeled in the subscripts). All results are shown in the dc limit ($\omega \to 0$).
}
\centering
    \renewcommand{\arraystretch}{2} 
    \setlength{\tabcolsep}{10pt}       
\begin{tabular}{c c  c}
\hline
\hline
Tensors &  Fermi surface & Fermi sea
\\ 
\hline
\hline
$\tilde{\chi}^{\gamma;a}_{\rm odd}$ & 0 & $-\dfrac{\mathcal{U}^{\gamma;a}_{np}}{2\hbar \omega^2_{np}}\tilde{\varepsilon}_{n\kb}f_n $ \\
$\tilde{\chi}^{\gamma;a}_{\rm even}$ & $\dfrac{\tau}{2}\tilde{\varepsilon}_{n\mathbf{k}}s^\gamma_{nn}v^a_{nn}\dfrac{\partial f_n}{\partial \varepsilon_{n \mathbf{k}}}$ & 0 \\[2ex]
\hline
$\tilde{\sigma}^{\gamma;ab}_{\rm odd}$ & $\dfrac{\tau}{2}\{v^a,s^{\gamma}\}_{nn}v^b_{nn}\tilde{\varepsilon}_{n\mathbf{k}}\dfrac{\partial f_n}{\partial \varepsilon_{n\mathbf{k}}}$ & 0 \\
$\tilde{\sigma}^{\gamma;ab}_{\rm even}$ & 0 & $\hbar~\Omega^{\gamma;ab}_{np}\tilde{\varepsilon}_{n\mathbf{k}} f_n $ \\
\hline
\hline
\end{tabular}
\label{tab:summary_table2}
\end{table}

Analogous to TSM susceptibility, the thermal spin conductivity also decomposes naturally as $\sigma^{\gamma;ab}=\sigma^{\gamma;ab}_{\text{Surf}}+\sigma^{\gamma;ab}_{\text{Sea}}$. The intraband Fermi surface contribution to TSC is given by, 
\begin{equation}\label{eq:TSC_surf}
\sigma^{\gamma;ab}_{\text{Surf}}(\omega)=
\frac{1}{2}\sum_{n\mathbf k}
      \Bigl(v^{a}_{nn}s^{\gamma}_{nn}+\mathcal V^{\gamma;a}_{n}\Bigr)\,
      v^{b}_{nn}\,
      \tilde\varepsilon_{n\mathbf{k}}\,
      \frac{\partial f_n}{\partial\varepsilon_{n \mathbf{k}}}\,
      \frac{1}{1/\tau-i\omega}\; .
\end{equation}
This term depends explicitly on the band-resolved anomalous spin velocity 
$\mathcal V^{\gamma;a}_{n}=\sum_{p\neq n}\mathcal V^{\gamma;a}_{np}$ and scales linearly with $\tau$ in the dc limit. The interband Fermi sea contribution is governed by the \emph{spin-geometric tensor} $T^{\gamma; ab}_{pn}$~\cite{spin_berry_curvature3}, 
\begin{equation}\label{eq:TSC_sea}
\sigma^{\gamma;ab}_{\text{Sea}}(\omega)=
 \hbar\sum_{n\neq p,\mathbf k}
      T^{\gamma; ab}_{pn}\,
      (\tilde\varepsilon_{n\mathbf{k}}f_{n}-\tilde\varepsilon_{p\mathbf{k}}f_{p})\,
      \frac{\omega_{np}}{1/\tau-i\!\bigl[\omega-\omega_{np}\bigr]} \; .
\end{equation}
The spin geometric tensor generalizes the usual quantum-geometric tensor to spin transport and is given by,  
\begin{equation}
T^{\gamma; ab}_{pn} = \frac{\{\hat{v}^a, \hat{s}^{\gamma}\}_{pn} v^b_{np}}{4 \varepsilon_{pn}^2} = \mathcal{G}^{\gamma; ab}_{pn} - \frac{i}{2} \Omega^{\gamma; ab}_{pn}.
\label{spin-quantum-metric-tensor}
\end{equation}
Here, $\mathcal G$ is the \emph{spin quantum metric}~\cite{spin_berry_curvature3} and $\Omega$ is the \emph{spin Berry curvature}~\cite{spin_berry_curvature1, spin_berry_curvature2, spin_berry_curvature3}. 

These results establish a unified framework for finite-frequency thermal spin responses of Bloch electrons rooted in spin band geometry. The framework applies broadly across materials and symmetry classes, including insulators and nonmagnetic systems, where such responses have so far remained largely unexplored. 

\textcolor{blue}{\it Symmetry constraints:--}
With the general formalism of TSM and TSC in place, we next analyze how fundamental and crystalline symmetries constrain these response tensors. 
This is crucial for identifying material platforms where TSM and TSC can manifest. From Eqs.~(\ref{eq:TSM_susc}, \ref{eq:TSC_surf}, \& \ref{eq:TSC_sea}), it is clear that TSM and TSC originate from the band geometric quantities ${\mathcal V}$, ${\mathcal U}$, ${\mathcal G}$, and ${\Omega}$. These quantities display distinct transformation properties under fundamental symmetries: \(\mathcal V^{\gamma;a}_{np}\) is even under time reversal but odd under inversion, whereas \(\mathcal U^{\gamma;a}_{np}\) is odd under both operations. Similarly, $\mathcal{G}^{\gamma; ab}_{pn}$ is even under inversion but changes sign under time reversal, whereas $\Omega^{\gamma; ab}_{pn}$ is even under both inversion and time reversal operations (see Sec.~\textcolor{blue}{S2} of the SM for more details~\cite{SM}). These contrasting symmetry properties ensure that TSM and TSC inherit both $\mathcal{T}$-odd and $\mathcal{T}$-even components.
The $\cal T$-odd responses vanish in non-magnetic systems, while $\cal T$-even components remain finite in both magnetic and non-magnetic systems. In Sec.~\textcolor{blue}{S3} of SM~\cite{SM}, we provide the explicit expressions for the ${\mathcal T}$-even and ${\mathcal T}$-odd components of finite frequency TSM and TSC responses. Table~\ref{tab:summary_table2} lists the ${\mathcal T}$-even and ${\mathcal T}$-odd components for both responses in the \textit{d.c.} limit. This decomposition in Table~\ref{tab:summary_table2} reveals a thermodynamic surprise. ${\cal T}$-odd TSM and ${\cal T}$-even TSC are intrinsic thermal spin responses, which arise purely from the interplay of the Fermi sea and spin band geometry, analogous to the anomalous Hall effect for charge flow. 

To analyze the impact of crystalline symmetries, we follow Neumann’s principle. It states that if a crystal is invariant under a certain set of symmetry operations, its physical response tensors must also remain invariant under the same operations. A general response tensor $\mathcal{P}^{ijk...}$ transforms under a symmetry operation $\mathcal{O}$ as~\cite{newnham2004,Briss_ROP1963,zhang_NSR2023,Brandmuller_CMA1986,Authier_book2013,Litvin_mpg2013,Gallego_2019} 
\begin{equation}
\mathcal{P}^{i'j'k'...} = \eta_{\mathcal{T}} \eta_{\mathcal{R}} \mathcal{O}^{i'i} \mathcal{O}^{j'j} \mathcal{O}^{k'k} \cdots \mathcal{P}^{ijk...},
\end{equation}
where $i,j,k,\dots$ are spatial indices. The factor $\eta_{\mathcal{R}}$ equals $+1$ for polar tensors and $\det(\mathcal O)$ for axial tensors.  The factor $\eta_{\mathcal{T}}$ captures the effect of time reversal. $\eta_{\cal T}=-1$ for $\mathcal{T}$-odd tensors under operations $\cal O$ involving $\mathcal{T}$ and $\eta_{\cal T}=+1$ otherwise~\cite{Gallego_2019}. The symmetry-imposed condition $\mathcal{P}^{i'j'k'...}=\mathcal{P}^{ijk...}$ ensures that only certain tensor components remain nonzero under a given symmetry operation $\mathcal O$. In our case, TSM response $\chi^{\gamma;a}$ is a rank-2 axial tensor while TSC response $\sigma^{\gamma;ab}$ is a rank-3 axial tensor, since both carry an odd number of axial indices. Further details of the symmetry analysis are presented in Sec.~\textcolor{blue}{S3} of the SM~\cite{SM}. 

In Table~\ref{symmetry_table}, we summarize the symmetry-allowed independent components of $\chi^{\gamma;a}$ and $\sigma^{\gamma;ab}$ across the 32 non-magnetic crystallographic point groups, assuming a thermal gradient in the $xy$ plane. This systematic classification in Table~\ref{symmetry_table} and extended in Sec.~S3 of the SM~\cite{SM} to include finite-frequency responses across all magnetic point groups, lays the foundation for a systematic search of materials hosting these responses. 

To substantiate our theoretical predictions in realistic and experimentally accessible materials, we investigate $\mathcal{T}$-even thermal spin responses in non-magnetic RhGe and $\mathcal{T}$-odd responses in antiferromagnetic CuMnAs in the \textit{d.c.} field limit.

\begin{table}[t!]
    \caption{{\bf Classification of thermal spin responses:} 
    List of symmetry-allowed TSM and TSC responses for all 32 non-magnetic crystallographic point groups, assuming a thermal gradient in the $xy$ plane. In non-magnetic systems, only ${\cal T}$-even responses are allowed, which correspond to an extrinsic Fermi surface contribution to TSM, and an intrinsic Fermi sea component in TSC.}
    \renewcommand{\arraystretch}{1.5}
    \setlength\tabcolsep{0.2 cm}
    \begin{tabular}{c| c}
    \hline
    \hline
    Tensor Components & Crystalline Point Groups \\
    \hline
    \hline
    $ \chi^{x;x},~\chi^{y;y}$ & $2.1,222,4,$-$4,422,-42m,3,32,6,$ \\
    & $622,23,432$\\
   \hline
    $\chi^{x;y}$,~$ \chi^{y;x}$ & $m,mm2,4,-4,4mm,3,3m,6,6mm$  \\
    \hline
    $ \chi^{z;x}$ & $2.1$ \\
    \hline
    $ \chi^{z;y}$ & $m$ \\
    \hline
    \hline
    $\sigma^{x;xx},~\sigma^{x;yy},~\sigma^{y;xy},$ &{$1,$-$1,3,$-$3,32,3m,$-$3m,$} \\
    $\sigma^{y;yx}$ & \\
    \hline
    $\sigma^{x;xy},~\sigma^{x;yx},\sigma^{y;xx},$ & $1,$-$1,2,m,2/m,3,$-$3,$ \\
    $\sigma^{y;yy}$ & \\
    \hline
    $\sigma^{z;xx},~\sigma^{z;yy},\sigma^{x;zx},$ & $1,$-$1,4,$-$4,4/m,3,$-$3,6,$-$6,6/m$ \\
    $\sigma^{y;zy}$ & \\
    \hline
    $\sigma^{z;xy},~\sigma^{z;yx},\sigma^{x;zy},$ & All 32 crystalline \\
    $\sigma^{y;zx}$ & point groups \\
    \hline
    \hline
    \end{tabular}
    \label{symmetry_table}
\end{table}

{\it \textcolor{blue}{TSM and TSC in {$\mathcal T$-symmetric} RhGe:-}}
\begin{figure*}
    \centering
    \includegraphics[width=0.8\linewidth]{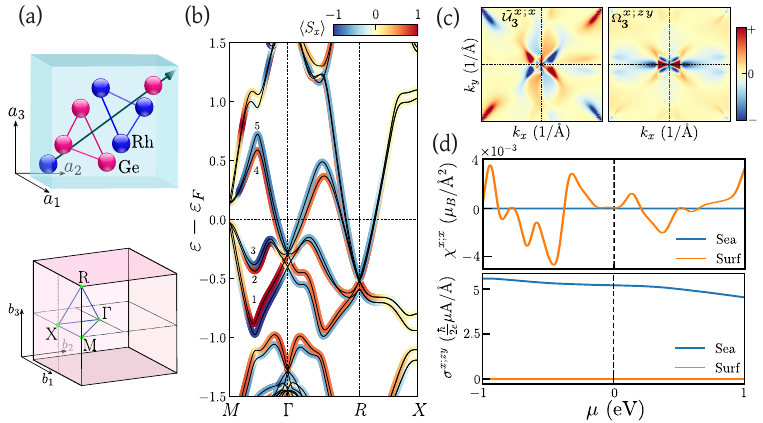}
    \caption{\textbf{Thermally driven spin magnetization and spin current in RhGe.} (a) Crystal structure of RhGe distorted along the $[111]$ axis (top) and the bulk Brillouin zone (bottom). (b) Bulk band structure with projected expectation value of $\hat{S}^x$. (c) Momentum-space distribution of the modified anomalous spin polarizability ($\tilde{\mathcal U}_n^{\gamma;a} = \sum_{p \ne n} {\mathcal U^{\gamma;a}_{np}}/\varepsilon_{np}^2$), and spin Berry curvature for the third energy band. (d) Dependence of the thermal spin magnetization and thermal spin conductivity components on chemical potential. In both responses, only the ${\cal T}$-even contributions are finite, which correspond to extrinsic Fermi surface response in magnetization, and intrinsic Fermi sea response in current.}
    \label{fig_2}
\end{figure*}
We show time-reversal symmetric thermal spin responses in the non-magnetic Kramers–Weyl metal RhGe using first-principles calculations. RhGe crystallizes in a B20-type chiral cubic structure [Fig.~\ref{fig_2}(a)], corresponding to the non-centrosymmetric space group P2$_1$3 (point group 23)~\cite{kamaeva2022structural,tsvyashchenko2016}. Sublattice displacements in RhGe along the [111] direction dimerize atomic bonds and break inversion and roto-inversion symmetries while preserving threefold rotational symmetry ($\mathcal{C}_3$) along the same axis. Our symmetry classification in Table~\ref{symmetry_table} shows that RhGe permits finite TSM response components $\chi^{x;x}$ and $\chi^{y;y}$, along with TSC response components $\sigma^{z;xy}$, $\sigma^{z;yx}$, $\sigma^{x;zy}$, and $\sigma^{y;zx}$. The electronic band structure,  shown in Fig.~\ref{fig_2}(b), agrees with previous report~\cite{sougata2024unconventional} (see Sec.~\textcolor{blue}{S4} of the SM for computational details \cite{SM}). The presence of time-reversal symmetry in RhGe enforces double degeneracy of bands at high-symmetry points such as $\Gamma$ and $R$, which act as hotspots for band-geometric quantities. This is evident in Fig.~\ref{fig_2}(c), which shows the momentum-space distribution of anomalous spin polarizability ($\mathcal{U}^{x;x}$) and spin Berry curvature ($\Omega^{x;zy}$) for the third band in the $k_z=0$ plane around the $\Gamma$ point.

Figure~\ref{fig_2}(d) shows the chemical potential ($\mu$) dependence of finite ${\cal T}$-even responses $\chi^{x;x}$ and $\sigma^{x;zy}$ at $T = 100$~K. As $\mathcal{U}^{x;x}$ is $\cal{T}$-odd, the corresponding Fermi-sea response vanishes in ${\cal T}$-symmetric RhGe. In contrast, $\Omega^{x;zy}$, being a $\cal{T}$-even quantity, generates a finite Fermi sea contribution to TSC. Quantitatively, the bulk TSM susceptibility reaches $\sim 10^{-2}\mu_B/\rm \AA^2$, while TSC conductivity can be as large as $\sim 5~\hbar/2e~\mu\rm A/\AA$. 
These results establish non-magnetic RhGe as a promising material for realizing sizable ${\cal T}$-even TSM and TSC responses.

\begin{figure*}
    \centering
    \includegraphics[width= 0.8\linewidth]{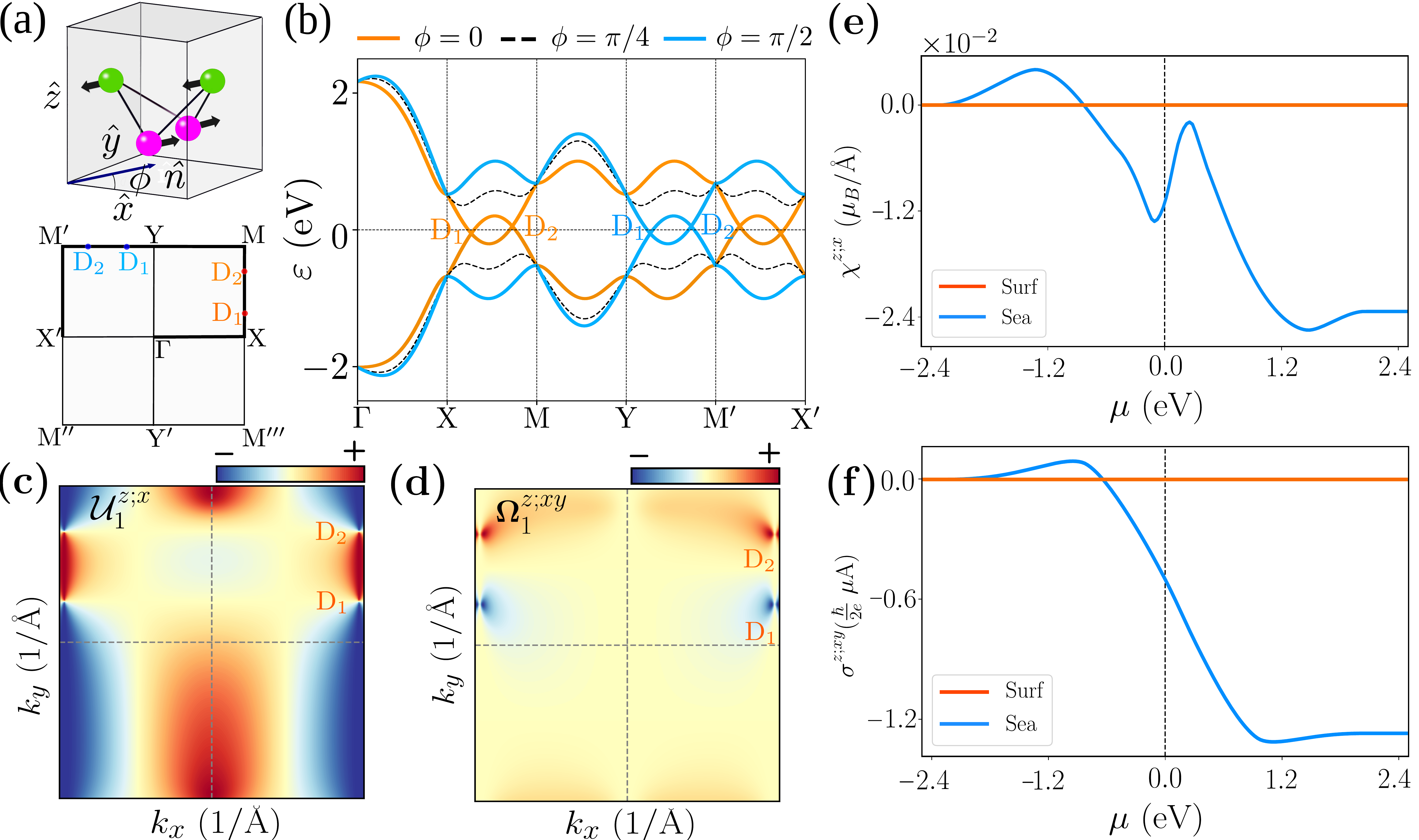}
    \caption{\textbf{TSM and TSC in topological antiferromagnet CuMnAs:} (a) Unit cell and rectangular Brillouin zone of CuMnAs. The green and purple spheres denote magnetic Mn atoms on opposite sublattices of the AFM. (b) The band structure of one layer of CuMnAs for different in-plane orientations of N\'eel vector, defined by the azimuthal angle $\phi$ with respect to the $\hat{x}$ axis. Dirac crossings appear only for $\phi=0\degree$ and $90\degree$. (c, d) $\mathbf{k}$-space distribution of band resolved anomalous spin polarizability $\mathcal{U}^{z;x}$ and spin-Berry curvature $\Omega^{z;xy}$ for the first band at $\phi=0$, 
    showing strong enhancement of the band geometric quantities near the band edges. (e, f) Chemical potential dependence of TSM tensor $\chi^{z;x}$ and TSC tensor $\sigma^{z;xy}$ at $T=50$ K.}
    \label{fig_3}
\end{figure*}

\textcolor{blue}{\it TSM and TSC in ${\cal T}$-broken CuMnAs:-}
To further demonstrate sizable TSM and TSC responses in time-reversal broken (${\cal T}$-odd) systems, we consider the topological antiferromagnet {CuMnAs}. In CuMnAs, collinear magnetic sublattices connected by $\cal PT$ symmetry cancel the net atomic magnetization, so the thermally induced itinerant spin magnetization dominates the response. We employ a minimal model of tetragonal {CuMnAs} on a crinkled quasi-2D square lattice with a collinear antiferromagnetic state~\cite{smejkal_prl17, Watanabe_prx21}, detailed in Sec.~\textcolor{blue}{S5} of the SM~\cite{SM}. The Néel vector lies in the $xy$ plane, as shown in Fig.~\ref{fig_3}(a). Tetragonal {CuMnAs} belongs to the point group $mmm'$ (same as $m'mm$ with crystallographic axes rotated), which hosts the combined $\cal PT$, ${\cal C}_{2}^z$, ${\cal M}_x$, and ${\cal M}_y$ symmetries, while breaking individual $\cal P$ and $\cal T$ symmetries. Figure~\ref{fig_3}(b) shows the band structure (doubly degenerate due to $\cal PT$ symmetry) for different orientations of the Néel vector in the $xy$ plane. For $\phi=0$ and $\pi/2$, Dirac cones form along the $\rm XM$ and $\rm YM'$ paths of the Brillouin zone (BZ). Figures~\ref{fig_3}(c) and (d) highlight enhanced band-geometric quantities near these Dirac points (for $\phi=0$). Figures~\ref{fig_3}(e) and (f) demonstrate the variation of the TSM tensor $\chi^{z;x}$ and TSC tensor $\sigma^{z;xy}$ with chemical potential $\mu$ at $T = 50$~K. The plots indicate significant 2D TSM susceptibility ($\sim 10^{-2}~\mu_B/\rm \AA$) and TSC conductivity ($\sim 1~\hbar/2e~\mu \rm A$). The underlying $\cal PT$ symmetry of CuMnAs ensures that only Fermi-sea contributions ($\cal PT$-even) remain finite, which is also evident from the results.  

Together, these findings demonstrate that both ${\cal T}$-symmetric RhGe and ${\cal T}$-broken CuMnAs support significant TSM and TSC responses, establishing them as good candidates for thermo-spintronic applications. Furthermore, our analysis of a magnetized Rashba 2DEG (see Sec~\textcolor{blue}{S6} of the SM~\cite{SM}) confirms that spin band geometry can produce sizable TSM and TSC responses near band edges, even in simple two-band systems.

\textcolor{blue}{\it Experimental implications:--} 
Our predictions for band-geometry-driven TSM and TSC are experimentally accessible. In RhGe, the intrinsic TSC can be detected via the inverse spin Hall effect (ISHE) in a RhGe/Pt bilayer, where a temperature gradient along the $y$ drives an $x$-polarized spin current flowing along $z$. For a realistic thermal gradient of $\nabla_y T = 1~{\rm K}/\mu m$ at $T = 100~\rm K$, readily achievable via Peltier elements or focused laser heating~\cite{uchida_2008, Rehm_pra2021}, we estimate sizable intrinsic spin current density $J^{x;z} \sim 5.5 \times 10^8~\hbar/2e~\rm A/m^2$ [see Fig.~\ref{fig_2}(d)]~\cite{jain_2023}. Injected into a heavy metal such as Pt, the spin current produces an electric field via ISHE, which can be measured using lock-in techniques~\cite{Sinova_2015, uchida_2008}. In a ferromagnet, the same spin current generates a spin torque corresponding to an effective field $B_{\rm eff}~\sim~0.1$ mT for permalloy layers~\cite{Krivorotov_science2005} of thickness $l\sim 1$ nm\footnote{The torque per unit magnetization is defined as $|\Gamma^s| = \gamma J^{s_x}_x /(M_s l)$, where $\gamma$ is the gyromagnetic ratio, $M_s$ the saturation magnetization, and $l$ the magnetic layer thickness~\cite{Sinova_rmp15, Liu_prl2011}. This torque results in an effective field, $B_{\rm eff}=\Gamma^s/\gamma = 0.1~\rm mT$, acting on localized magnetic moments in a permalloy-based ferromagnet (FM) having $M_s \sim 6.4 \times 10^5~{\rm A/m}$~\cite{Krivorotov_science2005} of width $l\sim 1~\rm nm$.}. In parallel, the TSM response can be probed optically via the polar Kerr effect~\cite{Stamm_prl2017, Ortiz_prb2022, Choi_AS2018}, particularly for energies where the spin susceptibility $\chi^{x;x}$ exhibits sharp features corresponding to band crossings [Fig.~\ref{fig_2}(c)]. With the same amount of thermal field, RhGe generates a peak spin magnetization of $\delta \mathcal{S}^x\approx 0.4\times 10^{-5}~\mu_B/\rm nm^3$. 

For CuMnAs, the TSC reaches $J^{z;x}\sim 1.2\times 10^{-2}~\hbar/2e~\rm A/m$ [see Fig.~\ref{fig_3}(f)] under $\mathbf{E}_T=0.01~\rm \mu m^{-1}~\hat{x}$. For the lattice constant of $c=6.32$\AA~\cite{Maca_prb2017, Wadley_NC2013}, this corresponds to a 3D current density of $J^{z;x}\sim 0.19\times 10^{8}~\hbar/2e~\rm A/m^2$ and is comparable to the TSC in RhGe. The 2D spin susceptibility peaks at  $\chi^{z;x}\sim 2.4\times10^{-2}~\mu_B/$\AA. This corresponds to a spin magnetization of $\delta \mathcal{S}^z\sim 2.4\times10^{-6}~\mu_B/\rm nm^2$ which is equivalent to a 3D magnetization of $\delta \mathcal{S}^z\sim 0.38\times 10^{-5}~\mu_B/\rm nm^3$. These estimates demonstrate that thermal spin responses are systematically enhanced near Dirac crossings and band-geometric hotspots, providing an experimentally accessible probe of spin quantum geometric contributions to thermal spin responses. 

%

\textcolor{blue}{{\it Conclusion:--}} 
Our results establish a unified spin band-geometric framework for TSM and TSC responses in itinerant electronic systems, addressing a long-standing gap in the field. We identify the spin-velocity metric tensor and spin geometric tensor as the key band-geometric quantities governing these responses, providing a direct analogue to Berry curvature and quantum metric in charge transport. This framework reveals that thermal spin responses can be sizable even in non-magnetic systems, broadening spin-caloritronic functionalities beyond conventional magnets. 

Our first-principles and model calculations on RhGe and CuMnAs illustrate how spin band-geometric quantities drive sizable thermal spin responses. In RhGe, symmetry-allowed tensor components enhance TSM and TSC near band crossings, while in CuMnAs, the $\cal PT$ symmetry only allows Fermi-sea terms. Our systematic symmetry classification of response tensors provides a roadmap for discovering new material platforms. These insights deepen the understanding of thermal spin transport and open new opportunities. Thermally driven spin signals, detectable optically or electrically, can be harnessed for switching, memory, and information processing in nanoscale devices.


\textcolor{blue}{{\it Acknowledgments:--}} 
We acknowledge the help received from Sougata Mardanya and Nirmalya Jana (IIT Kanpur) regarding ab initio and Wannier-based calculations. Sankar and Sayan acknowledge IIT Kanpur for funding support. Harsh acknowledges the Prime Minister's Research Fellowship under the Ministry of Education, Government of India for financial support. A.A. acknowledges funding from the Core Research Grant by ANRF (Sanction No. CRG/2023/007003), Department of Science and Technology, India. 

\bibliography{references}

\end{document}